\documentclass[aps,prl,twocolumn]{revtex4}%
\usepackage{graphicx}
\usepackage{amssymb}
\usepackage{graphicx}
\usepackage{amsmath}
\usepackage{amsfonts}
\usepackage{color}%
\providecommand{\U}[1]{\protect\rule{.1in}{.1in}}
\providecommand{\U}[1]{\protect\rule{.1in}{.1in}}

\begin{document}
\title{Tracking the time-evolution of the electron distribution function in Copper by
femtosecond broadband optical spectroscopy}
\author{Manuel Obergfell and Jure Demsar}
\affiliation{Institute of Physics, Johannes Gutenberg-University Mainz, 51099 Mainz, Germany}


\begin{abstract}
Multi-temperature models are nowadays often used to quantify the ultrafast
electron-phonon (boson) relaxations and coupling strengths in advanced quantum
solids. To test their applicability we study the time evolution of the
electron distribution function, f(E), in Cu over large range of excitation
densities using broadband time-resolved optical spectroscopy. Following
intraband optical excitation, f(E) is found to be athermal over several 100
fs, while substantial part of the absorbed energy already being transferred to
the lattice. We show, however, that the electron-phonon coupling constant can
still be obtained using the two-temperature model analysis, provided that the
data are analyzed over the time-window, when the electrons are already quasi
thermal, and the electronic temperature is determined experimentally.

\end{abstract}
\maketitle

Cooperative phenomena in quantum solids arise from a delicate balance among
interactions between charge, spin and lattice degrees of freedom. The
knowledge of the interaction strengths between the different subsystems is
thus crucial for their understanding. The knowledge of the Eliashberg
electron-boson coupling constant $\lambda$ is of particular interest in novel
superconductors, as it provides information on the significance of the
electron-boson interaction (or the lack thereof) for superconducting pairing.
One of the promising approaches to determine $\lambda$ is to use femtosecond
(fs) time-resolved techniques \cite{AllenEx}. Here, fs optical pulses are used
to excite the electronic system, while the recovery dynamics is probed by
measuring the resulting transient changes in optical constants
\cite{BrorsonEx} or the electronic occupation near the Fermi energy
\cite{Perfetti,Cnano}. Considering simple metals and assuming the
electron-electron ($e-e$) thermalization being much faster than the
electron-phonon ($e-ph$) relaxation, the so called two-temperature model (TTM)
has been put forward \cite{Kaganov,AllenEx}. Within this description, the
electrons rapidly thermalize to a temperature $T_{e}$, which can be much
higher than that of the lattice, $T_{l}$. This process is followed by the
$e-ph$ thermalization on a timescale $\tau_{e-ph}$, which is inversely
proportional to the $e-ph$ coupling strength. This widely used model suggests
a particularly simple relationship between the measured relaxation time and
$\lambda$, when experiments are performed at $T_{l}\gtrsim\Theta_{D}$;
$\Theta_{D}$ being the Debye temperature. With the electronic specific heat
$C_{e}=\gamma T_{e}$, where $\gamma$ is the Sommerfeld constant, the time
evolutions of $T_{e}$ and $T_{l}$ are given by a set of coupled heat equations
\cite{AllenEx,BookChapter}. Here $\frac{\partial T_{e}}{\partial t}%
=(T_{l}-T_{e})/\tau_{e-ph}$, with \cite{AllenEx}
\begin{equation}
\frac{1}{\tau_{e-ph}}=\frac{3\hbar\lambda\left\langle \omega^{2}\right\rangle
}{\pi k_{B}T_{e}}(1-\frac{\hbar^{2}\left\langle \omega^{4}\right\rangle
}{12\left\langle \omega^{2}\right\rangle k_{B}^{2}T_{e}T_{l}}%
+..),\label{gamma}%
\end{equation}
where $\lambda\left\langle \omega^{n}\right\rangle =2%
{\textstyle\int\nolimits_{0}^{\infty}}
\left[  \alpha^{2}F\left(  \Omega\right)  /\Omega\right]  \Omega^{n}\,d\Omega$
while $\alpha^{2}F\left(  \Omega\right)  $ is the product of the $e-ph$
coupling strength $\alpha^{2}$, and the phonon density of states $F$. Often
the Debye approximation is used, where the coupling strength is mode
independent. In this case $\left\langle \omega^{2}\right\rangle $ is the mean
square phonon frequency.

Following pioneering works on noble metals
\cite{Eesley83,Fujimoto,Schoenlein,Elsayed,ElsayedAliIntDep,Sun} numerous
time-resolved experiments\ on superconductors have been performed, ranging
from conventional \cite{BrorsonEx}, to high-T$_{c}$ cuprate
\cite{Perfetti,Brorson2E,ChekalinE,Gadermaier,DalConte,Chia2013} and pnictide
\cite{MansartPnic,Stojchevska} superconductors, aiming at the determination of
$\lambda$. Similar studies were performed also on other advanced materials
ranging from carbon nanotubes \cite{Cnano,Cnanotubes}, ferromagnets
\cite{CarpeneFe}, to metallic nanoparticles \cite{Voisin}. Moreover, several
extensions to multi-temperature models have also been proposed to account for
experimental observations \cite{DalConte,Chia2013}. Despite the reasonable
agreement between the experimentally extracted and theoretically estimated
values of $\lambda\left\langle \omega^{2}\right\rangle $ \cite{BrorsonEx}
numerous studies shed doubts on the applicability of the TTM. The
time-resolved photoemission data on Au showed that even at room temperature
and high excitation densities, the $e-e$ thermalization time is as long as 800
fs while the electronic distribution at earlier times is strongly athermal
\cite{Fann}. The second major shortcoming of the TTM is the prediction that in
the limit of weak excitations $\tau_{e-ph}\varpropto T_{l}^{-3}$ as
$T_{l}\rightarrow0$ K, which was never observed in simple metals
\cite{Groenevald,Ahn}. Both, the absence of slowing down of relaxation at
low-$T_{l}$ and long $e-e$ thermalization times were attributed to Pauli
blocking, where $e-e$ scattering into states below the Fermi level ($E_{F}$)
is reduced due to the small fraction of unoccupied states to which electrons
can be scattered to \cite{Groenevald,Ahn}. Moreover, several recent studies of
dynamics in advanced solids suggest that the $e-e$ and $e-ph$ thermalization
timescales are actually comparable \cite{Kusar,Pashkin,Beyer}.

In this Letter we present an all-optical approach to study the time-evolution
of the photoinduced changes in the electronic distribution function near the
Fermi energy, $\Delta f(E-E_{F})$, in thin copper films. We achieve that by
studying the temporal evolution of the complex dielectric function,
$\varepsilon(\omega)=\varepsilon_{1}(\omega)+i\varepsilon_{2}(\omega)$,
following intraband photoexcitation. In Cu $\varepsilon(\omega)$ is in the
visible spectral range largely governed by the interband transition from the
d-band to the Fermi level. With the combination of static $\varepsilon
(\omega)$, thermomodulation, $\Delta\varepsilon(\omega)=$ $\frac
{d\varepsilon(\omega)}{dT}\Delta T$, and the simple model of the electronic
density of states, which accounts for $\varepsilon(\omega,T)$, we show that
$\Delta f(E,t)$ can be extracted from $\Delta\varepsilon(\omega,t)$. We show
that in Cu the $f(E)$ is quasi-thermal only for time delays larger $>0.5$ ps.
Moreover, the experimentally determined $T_{e}$'s are - for short time delays
- substantially lower than the expected values based on the absorbed energy
density. This implies a substantial energy transfer to the lattice already in
the early stage of relaxation. Despite the obvious disparity of the presented
results and the TTM, we demonstrate that the TTM can account for the data, yet
just for\ the time delays when the electron subsystem is already
quasi-thermal, and provided that $T_{e}$'s are determined experimentally, as
in our case. The presented approach could be generalized to other systems with
interband optical transitions in the visible range.

The broadband fs time-resolved optical studies were performed on thin (24 nm)
Cu films sputtered on (100) MgO substrate. The reflectivities and
transmissions of films were measured with commercial FTIR and UV-Vis
spectrometers. The measured optical constants were found to be in good
agreement with literature values \cite{Cudielectric}. The samples were
photoexcited by 50 fs near-infrared (NIR) pulses ($\lambda_{pe}=800$ nm, 1.55
eV). The absorbed energy densities, $U$, calculated from the measured
$\varepsilon(1.55$ eV$)$ \cite{SI} were varied between 4-250 J/cm$^{3}$. The
photoinduced changes of both transmission and reflectivity between 1.25 - 2.8
eV were measured with white light supercontinuum pulses generated in sapphire
\cite{Alfano}. Combining the static $\varepsilon(\omega)$ and the measured
transient changes in reflectivity and transmission, $\Delta\varepsilon
(\omega,t)$ is determined by numerically solving a system of appropriate
Fresnel equations \cite{SI,Dupertuis1,Dupertuis2}.

\begin{figure}[h]
\centerline{\includegraphics[width=90mm]{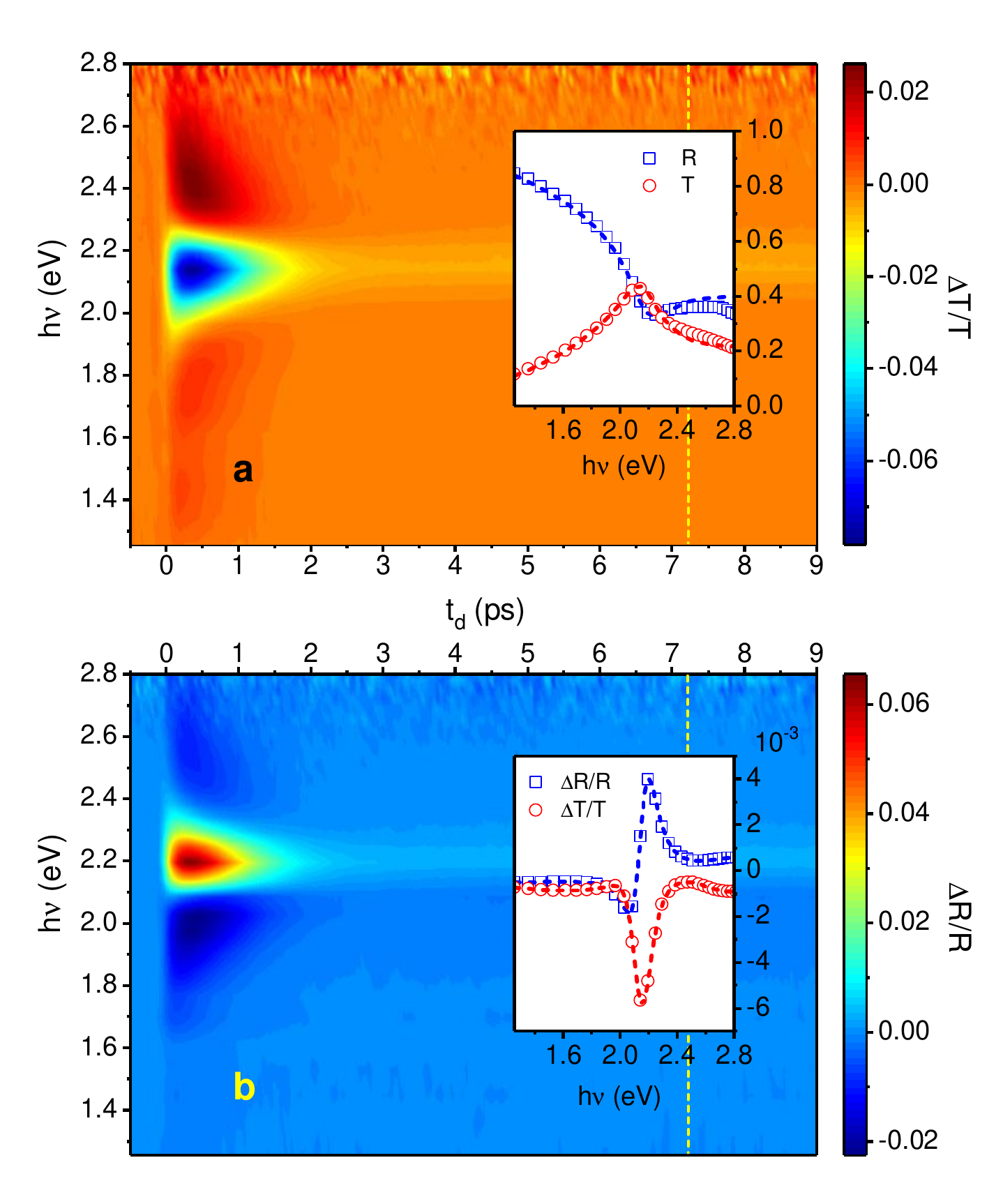}}\caption{Dynamics of the
photoinduced change in transmission (a) and reflectivity (b) of a 24 nm thick
Cu film on MgO substrate excited by a 50 fs NIR pulse. The base temperature is
300 K, and the excitation fluence is F = 3.4 mJ/cm$^{2}$, corresponding to the
absorbed energy density U = 54 J/cm$^{3}$. Inset to panel (a) presents the
equilibrium reflectivity (Re) and transmission (Tr) data (open symbols),
together with the corresponding model fits (dashed lines). Inset to panel (b)
presents the induced changes in Re and Tr at the time delay of 7.2 ps (open
symbols) together with the thermomodulation fit (dashed lines).}%
\end{figure}

Figure 1 shows the time evolution of the induced transmission, $\Delta$Tr/Tr,
and reflectivity, $\Delta$Re/Re, of a 24 nm thick Cu film on MgO substrate,
recorded at room temperature, in the spectral range between 1.25 and 2.8 eV.
The equilibrium reflectivity and transmission of the Cu thin film are
presented in insert to panel (a). The anomaly centered at $\approx2.1$ eV is a
result of the interband transition ($\mathcal{T}_{d-p}$) between the d-band,
located at $\mathcal{E}_{d-p}=2.1$ eV below the Fermi level, and the Cu s-p
band. The time-resolved data show strong changes in optical properties near
$\mathcal{E}_{d-p}$, arising from photoinduced changes in $\mathcal{T}_{d-p}$
\cite{Rosei}. Since the NIR pump-pulse excites the s-p band electrons, it is
the photoinduced Fermi-level smearing, i.e. broadening of the electronic
distribution near $E_{F}$, that is mainly responsible for changes in
$\mathcal{T}_{d-p}$ (see Figure 2a). Assuming the validity of the TTM, $T_{e}$
should reach $\approx1100$ K at U = 54 J/cm$^{3}$. Following the $e-e$ and
$e-ph$ thermalization processes, a quasi-equilibrium is reached within a few
picoseconds (the photoinduced spectra show no measurable changes between 5 and
30 ps), with the subsequent decay governed by the heat diffusion into the
substrate. Therefore, we can assume that $\Delta\varepsilon(\omega,t\gtrsim5$
ps$)=$ $\frac{d\varepsilon(\omega)}{dT}\Delta T$, where $\Delta T$ is the
resulting temperature increase, given by $U=%
{\textstyle\int\nolimits_{T_{0}}^{T_{0}+\Delta T}}
C_{p}\left(  T\right)  dT$, where $C_{p}\left(  T\right)  $ is the total
specific heat. Indeed, the recorded $\Delta$T/T($7$ ps) and $\Delta$R/R($7$
ps), shown in inset to Fig. 1(b), match well the changes obtained by simply
heating up the sample using a hot-plate (conventional thermomodulation). For U
= 54 J/cm$^{3}$ we obtain $\Delta T\approx15$ K (to be compared with the
estimated maximal $\Delta T_{e}\approx800$ K for early times).

\begin{figure}[h]
\centerline{\includegraphics[width=90mm]{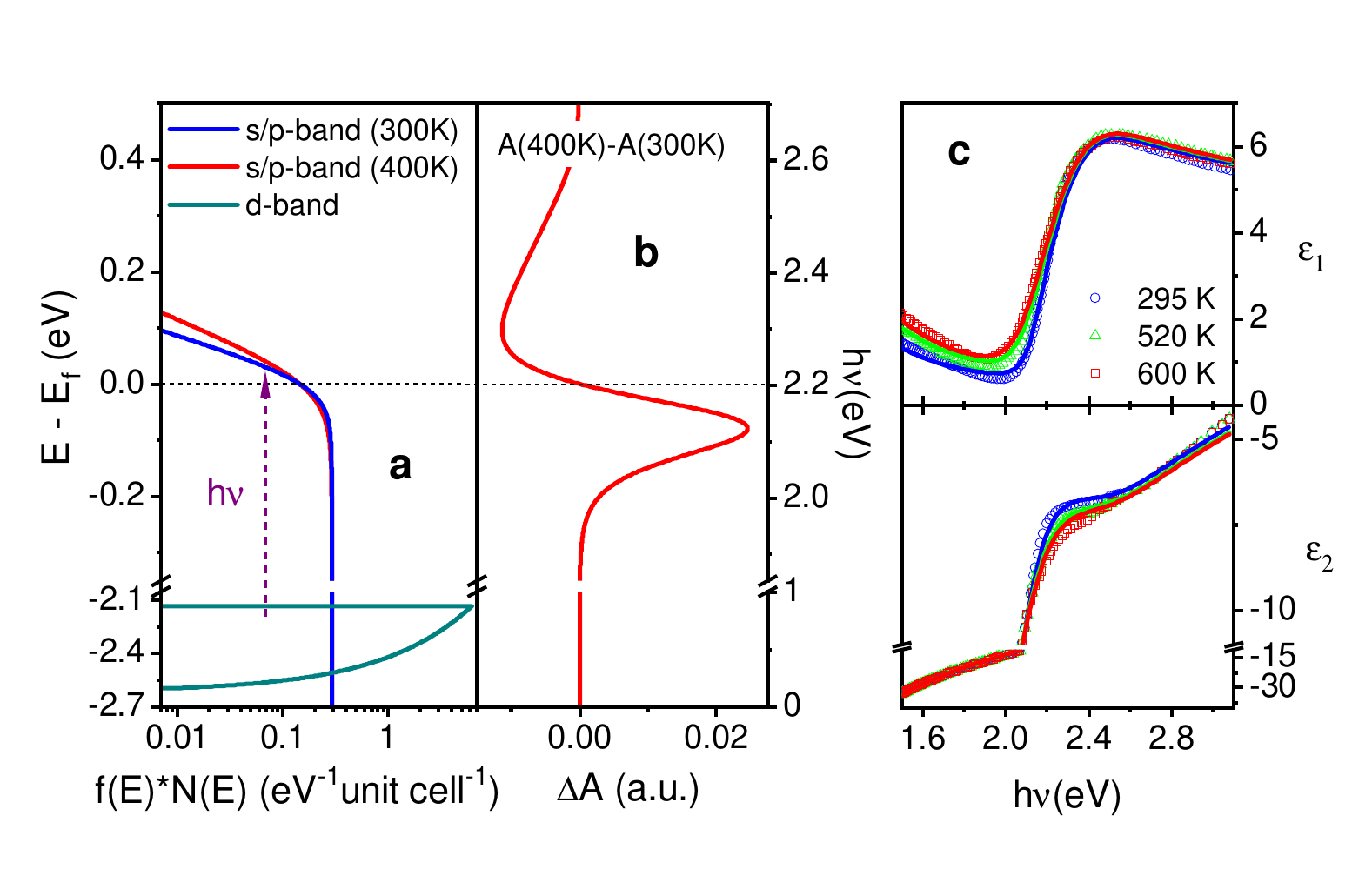}}\caption{The thermal
modulation of optical constants in Cu, modeled \ by a simplified model of the
density of states and Fermi golden rule. Panel (a) presents the modeled
occupied density of states in Cu (note the semi-logarithmic scale) at 300 and
400 K. The corresponding changes in absorption are presented in (b). These are
dominated by the changes in the interband transition between the fully
occupied d-band and the s-p conduction band. In panel (c) the model is applied
to fit (solid lines) the published\cite{Hanekamp} experimental spectroscopic
ellipsometry data on Cu taken at different temperatures (open symbols).}%
\end{figure}

As noted, the dominant contribution to changes in the optical constants in the
visible range stems from the photoinduced Fermi level smearing. The Fermi
level smearing results in opening/blocking the optical transitions from the
fully occupied $d$-band at $\approx2.1$ eV below E$_{F}$, to s-p band states
below/above the E$_{F}$ \cite{Rosei}, as sketched in Figure 2(a). Thus, a
proper parametrization of $\varepsilon(\omega,T)$ provides means to a direct
access to changes in the electronic distribution function - also for
time-delays where the distribution may be athermal.

To model the equilibrium and the thermomodulation optical spectra in the
visible range \cite{SI}, we consider $\varepsilon(\omega,T)=\varepsilon
_{D}(\omega,T)+\varepsilon_{d-p}(\omega,T)+\varepsilon_{\infty}(T)$. Here
$\varepsilon_{D}(\omega)$ is the free carrier Drude response of the s/p
electrons, $\varepsilon_{d-p}(\omega)$ describes the interband transition
between the uppermost d-band and the s-p band, while $\varepsilon_{\infty}$
sums up the contributions of higher energy interband transitions to
$\varepsilon(\omega,T)$. All contributions are temperature dependent. The
T-dependence of $\varepsilon_{D}(\omega)$ is governed by the T-dependence of
the Drude scattering rate, $\gamma_{D}$ \cite{SI}. For temperatures above 300
K, $\gamma_{D}$ is governed by the $e-ph$ scattering and thus depends linearly
on $T_{l}$. The changes due to the Fermi level smearing are, for the
thermomodulation, sketched in Figure 2b. They give rise to a bipolar change in
the interband absorption near $\mathcal{E}_{d-p}$, with the amplitude
proportional to $\Delta T_{e}$. In addition, a small shift of $\mathcal{E}%
_{d-p}$ can be expected, either due to the shift of the chemical potential
(proportional to $\Delta T_{e}$) or due to the thermal lattice expansion
(proportional to $\Delta T_{l}$). In Cu, the electronic DOS at E$_{F}$ is
nearly constant and the former can be neglected, thus the shift in
$\mathcal{E}_{d-p}$ is governed by $T_{l}$. Finally, the induced changes in
higher energy interband transitions (%
$>$
4 eV) may also contribute to $\Delta\varepsilon(\omega)$ in the visible range.
These changes, driven by the thermal expansion ($\propto\Delta T_{l}$) give
rise to a weak frequency-independent offset in the real part of $\Delta
\varepsilon(\omega)$ \cite{SI}. The findings are tested on published optical
data at different temperatures and presented in Figure 2c.

Since only $\Delta\varepsilon_{d-p}(\omega,T)$ is dominated by $\Delta
f(E,t)$, and $\Delta T_{l}\ $is much smaller than\ $\Delta T_{e}$, we can
parametrize the changes of $\varepsilon(\omega)$ that are a result of $\Delta
T_{l}$, thereby getting access to $\Delta f(E,t)$. We start by modeling the
equilibrium $\varepsilon(\omega,T)$, to account for Tr and Re at room
temperature, as well as for the bolometric responses. The latter is given by
$\Delta$Tr/Tr$(t\gtrsim5$ ps$)$ and $\Delta$Re/Re$(t\gtrsim5$ ps$)$ and was
recorded at 12 different excitation levels with $4<U<250$ J/cm$^{3}$. To model
$\varepsilon_{d-p}(\omega,T)$, which dominates $\Delta\varepsilon(\omega)$ in
the visible range, we developed a simple model (see \cite{SI}) considering the
Fermi golden rule, and using the band dispersions that give rise to densities
of states of the d-band and the s-p band, as shown in Fig. 2a. For the Drude
scattering rate, $\gamma_{D}$,\ and $\mathcal{E}_{d-p}$ we assume they depend
linearly on lattice temperature (e.g., for $\gamma_{D}$ we assume that
$\gamma_{D}(300K+\Delta T_{l})=\gamma_{300K}+c_{\gamma}\Delta T_{l}$). Such a
linear expansion is justified since the maximal changes in the lattice
temperature (for highest excitation densities used here) are of the order of
$\Delta T_{l}=60$ K. We determined these parameters by globally fitting the
equilibrium $\epsilon(\omega)$ and $\Delta\varepsilon(\omega,t\gtrsim5$ ps$)$
for $U$ spanning nearly two orders of magnitude. The resulting $\varepsilon
(\omega,T)$ is shown to describe well Tr and Re (inset to Fig. 1a) as well as
the bolometric (thermomodulation) response (inset to Fig. 1b).

To determine $\Delta f(E,t)$ from experimental data, we assume that $\Delta
T_{l}=\Delta T_{l}\left(  t\gtrsim5\text{ ps}\right)  \left[  1-\exp
(-t/\tau)\right]  $, where $\tau$ is the\ decay time of the spectrally
averaged transient. Both experimental studies \cite{Ralph} and detailed
numerical calculations \cite{Oppeneer} demonstrated that the phonon subsystem
is also athermal on the picosecond timescale. However, the relatively small
contribution of the components linked to changes in $T_{l}$ to the overall
changes in $\epsilon(\omega)$ make the result relatively insensitive to the
variation of $\tau$. With this, and the extracted coefficients describing
$\gamma_{D}(T_{l})$, $\varepsilon_{\infty}(T_{l})$, $\mathcal{E}_{d-p}(T_{l}%
)$, we obtain $\Delta f(E,t)$ by fitting the model to the experimental data.

\begin{figure}[h]
\centerline{\includegraphics[width=90mm]{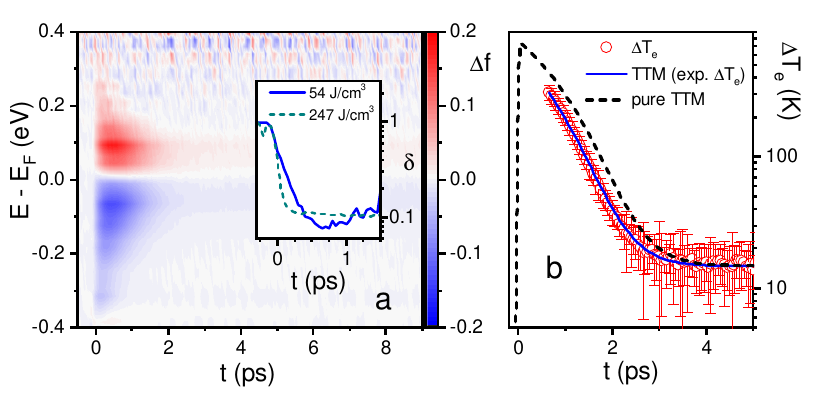}}\caption{Analysis of the
time-resolved optical data in Cu thin film ($U=54$ J/cm$^{3}$). Panel (a)
presents $\Delta f(E,t)$ following the photoexcitation. (b) The extracted
$\Delta T_{e}$ determined by fitting the data with the corresponding change of
the Fermi-Dirac distribution function, $\Delta f_{FD}$. The inset to panel (a)
presents the deviation between the measured $\Delta f(E,t)$ and $\Delta
f_{FD}(E,t)$ denoted by $\delta$ (see text), normalized to the absolute signal
strength. The results for two excitation densities are shown by the blue solid
line and the dashed green line. The minima of $\delta$ define the point where
electrons are well described by T$_{e}$. Panel (b) presents the fit to the
experimental $\Delta T_{e}$ (open red circles) using the TTM (solid blue
line). For comparison, we present the prediction of the pure TTM (dashed
line), where the initial electronic temperature has been calculated from $U$.
}%
\end{figure}

Figure 3 (a) presents the time evolution of $\Delta f$ extracted from the data
shown in Fig. 1. To evaluate $\Delta f(E,t)$ we first compare the experimental
$\Delta f$ around E$_{F}$ with the best fit assuming thermalized electrons,
where $\Delta f_{FD}=f(E,T_{e})-f(E,300$ K$)$, and T$_{e}$ is obtained by the
best fit to the experimental $\Delta f$ . The normalized error $\delta(t)=%
{\textstyle\sum}
\left\vert \Delta f(E,t)-\Delta f_{FD}(E,t)\right\vert /%
{\textstyle\sum}
\left\vert \Delta f_{FD}(E,t)\right\vert $, where the sum spans the data for
-0.4 eV $\lesssim E-E_{F}\lesssim$ 0.4 eV, is shown in inset to panel (a) for
two excitation densities. It follows that for U $\sim$ 50 J/cm$^{3}$ the
$f(E,t)$ reaches the quasi-thermal state only on the timescale of $\approx6$00
fs, while for U $\sim$ 250 J/cm$^{3}$ the electron thermalization time is
reduced to $\approx3$00 fs. Figure 3 (b) presents the time-evolution of the
extracted electronic temperature ($U=54$ J/cm$^{3}$) from the point, where
$f(E,t)$ is quasi-thermal. Analyzing experimental $\Delta T_{e}(t>0.6$ ps$)$
using the TTM (solid blue line) we obtain $\lambda\left\langle \omega
^{2}\right\rangle =45$ meV$^{2}$, which is in excellent agreement with
theoretical estimates \cite{Beaulac}. Importantly, the measured $\Delta T_{e}%
$'s are throughout the thermalization process substantially lower than
expected from the pure TTM, using the same value of $\lambda\left\langle
\omega^{2}\right\rangle $ and $\Delta T_{e,theo}(t=0)=\sqrt{T_{l}%
^{2}+2U/\gamma}-T_{l}$(dashed line in Figure 3b). This implies a substantial
energy transfer to the phonon subsystem already during the e-e thermalization process.

Not being able to properly determine $\Delta T_{e}$, as in most all-optical
studies, can be a major source of error in estimating $\lambda\left\langle
\omega^{2}\right\rangle $. Thus, the values of $\lambda\left\langle \omega
^{2}\right\rangle $ obtained by time-resolved optical methods vary
substantially. In Figure 4 we plot the extracted $\lambda\left\langle
\omega^{2}\right\rangle $ obtained at different excitation densities (black
spheres). As expected for moderate excitation densities \cite{Lin,Rethfeld1},
$\lambda\left\langle \omega^{2}\right\rangle $ is found to be independent of
$U$, provided that we use the experimentally extracted $\Delta T_{e}(t)$ for
the TTM analysis. Applying the common approach of extracting $\lambda
\left\langle \omega^{2}\right\rangle $ using $T_{e,theo}(t=0)$, either by i)
an exponential decay fit and Eq.(\ref{gamma}) or ii) by the full TTM fit, the
extracted $\lambda\left\langle \omega^{2}\right\rangle $ is shown to strongly
vary with $U$ (Figure 4a). Indeed, the previously published data on
$\lambda\left\langle \omega^{2}\right\rangle $ in Cu
\cite{BrorsonEx,ElsayedAliIntDep} seem to follow this trend.

\begin{figure}[h]
\centerline{\includegraphics[width=90mm]{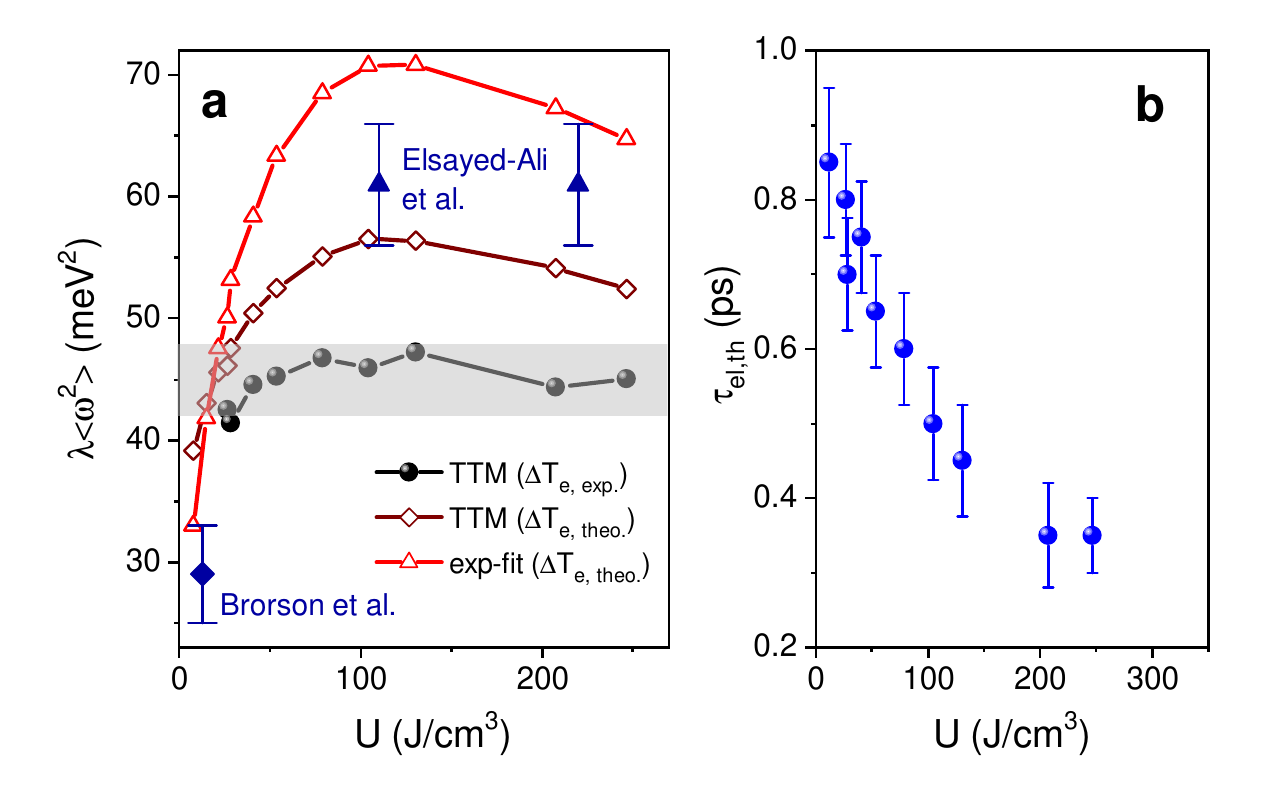}}\caption{(a) The
$\lambda\left\langle \omega^{2}\right\rangle $ extracted from the time
evolution of $\Delta T_{e}$ recorded at different excitation densities, $U$.
The value is independent on $U$, as expected for moderate excitation
densities. For comparison, we present $\lambda\left\langle \omega
^{2}\right\rangle (U)$ obtained by assuming the validity of the TTM ($\Delta
T_{e}^{theo}$), fit by either TTM or by a single exponential decay. For both,
the extracted $\lambda\left\langle \omega^{2}\right\rangle $ varies strongly
with $U$. The published values obtained from exponential fits
\cite{BrorsonEx,ElsayedAliIntDep} are included. Panel (b) presents the
excitation dependence of the electron-electron thermalization time,
$\tau_{et,th}$, obtained from the analysis of experimentally measured $\Delta
$f using the error analysis (see inset to Fig. 3a).}%
\end{figure}

Figure 4b presents the excitation dependence of the extracted electron
thermalization time, $\tau_{el,th}$. It shows that the main reason for the
departure of the observed relaxation dynamics from the standard TTM lies in
the slow $e-e$ thermalization at low excitation densities. The observation is
consistent with the relaxation of Pauli-blocking for high excitation
densities, and in-line with calculations using Boltzmann collision integrals
\cite{Rethfeld1}.

Our study demonstrates that, for simple metals with relatively weak e-ph
coupling, $\lambda\left\langle \omega^{2}\right\rangle $ can indeed be
extracted using the TTM, provided that the analysis is restricted to times
after the electronic distribution reached a quasi thermal one, and that the
electronic distribution function and $T_{e}$ is determined experimentally.
However, such studies should be performed as a function of excitation density
to test if the data are consistent with the TTM model predictions to begin
with (e.g. fluence dependent relaxation rate).

In numerous advanced solids, the (initial) carrier relaxation dynamics are
found to proceed on a sub-picosecond timescale, with the excitation dependent
studies showing fluence independent dynamics \cite{Gadermaier}. It has been
argued that, in the weak excitation limit, in many systems the $e-e$
thermalization times may actually be longer than the $e-ph$ relaxation times
over most of the accessible temperature range \cite{KabAlex,Baranov}. For such
a case, where the electronic distribution is athermal through most of the time
window of interest, an alternative analytic expression, linking the
experimentally measured $e-ph$ relaxation time $\tau_{e-ph}$ to the $e-ph$
coupling constant has been derived: $\lambda\left\langle \omega^{2}%
\right\rangle =\frac{2\pi k_{B}T_{l}}{3\hbar\tau_{e-ph}}$
\cite{KabAlex,Baranov}. Note, that the expression is very similar to the one
for the TTM, with the main difference being a factor of 2 and the $T_{l}$
instead of the $T_{e}$. In the low excitation density limit in Cu, the
relaxation time becomes independent on fluence \cite{SI}, and the extracted
$\tau_{el,th}$ (Fig. 4b) becomes comparable to the $\tau_{e-ph}$. Indeed, in
the low excitation density limit, we obtain $\lambda\left\langle \omega
^{2}\right\rangle =50$ meV$^{2}$ using the above expression of $\lambda
\left\langle \omega^{2}\right\rangle $ for the fully non-thermal case. The
value is close to the value extracted at high excitation densities using the
TTM. However, upon increasing the excitation density, the value of
$\lambda\left\langle \omega^{2}\right\rangle $ obtained from the non-thermal
model starts to decrease, signifying the change from the non-thermal to the
thermal regime.

As demonstrated above, the values of $\lambda\left\langle \omega
^{2}\right\rangle $ extracted from the time-resolved data may vary by as much
as a factor of 2, depending on the excitation density and the approximation
used. While this factor may appear to be small, we should note that this
difference may correspond to the difference between the strong and weak
coupling limits for superconductivity \cite{TheoryTc}. The demonstrated
all-optical approach, where the time evolution of the electronic distribution
function in a thin (bulk) film can be recorded, may provide a way to
unambiguously extracting $\lambda$'s also for advanced superconductors.

\begin{acknowledgments}
This work was supported by the Carl-Zeiss Stiftung and the DFG in the
framework of the Collaborative Research Centre SFB TRR 173 \textquotedblleft
Spin +X\textquotedblright. We gratefully acknowledge valuable discussions with
V.V. Kabanov and B. Rethfeld.
\end{acknowledgments}

\end{document}